\documentclass[%
 reprint,
superscriptaddress,
showpacs,
amsmath,amssymb,
aps,
]{revtex4-2}
\usepackage{graphicx}
\usepackage{dcolumn}
\usepackage{bm}
\usepackage[usenames]{color}
\usepackage{soul}
\usepackage{multibib}

\begin{document}

\preprint{APS/123-QED}

\title{Rapid fair sampling of XY spin Hamiltonian with a laser simulator}

\author{Vishwa Pal}\email{Corresponding author: vishwa.pal@iitrpr.ac.in}
\affiliation{Department of Physics, Indian Institute of Technology Ropar, Rupnagar 140001, Punjab, India}%
\affiliation{Department of Physics of Complex Systems, Weizmann Institute of Science, Rehovot 7610001, Israel}%
\author{Simon Mahler}%
\author{Chene Tradonsky}%
\author{Asher A. Friesem}%
\author{Nir Davidson}%
\affiliation{Department of Physics of Complex Systems, Weizmann Institute of Science, Rehovot 7610001, Israel}%

\date{\today}

\begin{abstract}
Coupled oscillators such as lasers, OPO's and BEC polaritons can rapidly and efficiently dissipate into a stable phase locked state that can be mapped onto the minimal energy (ground state) of classical spin Hamiltonians. However, for degenerate or near-degenerate ground state manifolds, statistical fair sampling is required to obtain a complete knowledge of the minimal energy state, which needs many repetitions of simulations under identical conditions. We show that with dissipatively coupled lasers such fair sampling can be achieved rapidly and accurately by exploiting the many longitudinal modes of each laser to form an ensemble of identical but independent simulators, acting in parallel. We fairly sampled the ground state manifold of square, triangular and Kagome lattices by measuring their coherence function identifying manifolds composed of a single, doubly degenerate, and highly degenerate ground states, respectively.  

\end{abstract}

\maketitle


Various combinatorial optimization problems that occur in social networks, neural networks, management of large data sets, artificial intelligence, spin glass, drug discovery, protein folding, traveling salesman etc, are considered to be computationally hard problems \cite{Steiglits98,Yamamoto17}. Such optimization problems can be mapped into classical spin systems (Ising or XY Hamiltonian), where they are reduced to finding the global minimum of the spin Hamiltonian \cite{Berloff17,Lagoudakis17,McMahon16,Takeda18}. There has been significant interest in building efficient simulators that are based on physical systems, and recently some have been realized. These include simulators that involve coupled lasers \cite{Nixon13, Vishwa17}, Bose-Einstein condensate (BEC) polaritons \cite{Berloff17} and optical parametric oscillators (OPOs) \cite{Yamamoto17,McMahon16, Inagaki16, Marandi14}. The success of this approach relies on finding efficiently and rapidly the ground state of the spin Hamiltonian \cite{Berloff17,Marandi14}. However, if the ground state is degenerate or nearly degenerate, this ground state manifold must be fairly sampled in order to obtain the full knowledge of the minimal energy state of the system, requiring many repetitions of the simulations under exactly the same conditions \cite{Tamate16, Takeda18, Konz19, Mandra17, Farhi}. 

In this letter, we present a new simulator for the XY spin Hamiltonian based on linearly coupled lasers that rapidly performs fair sampling by exploiting massive parallelism in the lasers spectral domain. Under the assumption of constant field amplitudes, the coupled lasers are well approximated as Kuramoto phase oscillators \cite{Acebron05}. Then the phases of the lasers can be mapped to the classical XY spins, and the ground state of the classical XY Hamiltonian can be analogous to phase locked steady state of the coupled lasers \cite{Nixon13}. Unlike finding the ground state of spin systems by cooling externally (related to the well-known Kibble-Zurek mechanism \cite{Zurek14, Mermin79, Coullet89, Kibble76, Zurek85,Navon15}), in coupled lasers the internal dissipation caused by coupling, drives the lasers into a globally stable phase locked state (minimal loss state), identical to the ground state of classical XY spin Hamiltonian \cite{Nixon13, Vishwa17}. The advantages of dissipative mechanisms were also demonstrated in OPO's and BEC polaritons simulators \cite{Berloff17, Yamamoto17, Marandi14}. 

In our simulator, each laser consists of approximately 250 longitudinal modes that form an ensemble of approximately 250 identical but independent simulators of the XY spin Hamiltonian. This provides a massive parallelism that enables rapid and accurate fair sampling of the ground state manifold. In earlier work, we used nonlinear coupling where all the longitudinal modes formed all together a single simulator \citep{Mahler19}. We directly measure the statistical average of spin ordering (magnetization) of the ground state manifold by measuring coherence between the lasers in different lattice geometries having single, double, and many degenerate ground states. 

The coherence between the lasers is described as \cite{Micha11}
\begin{equation}
V_{ij}=\sqrt{\langle \cos(\phi_{ij})\rangle^2 +\langle \sin(\phi_{ij})\rangle^2},
\end{equation}
where $\phi_{ij}=\phi_{i}-\phi_{j}$, and $\phi_{i}$ and $\phi_{j}$ are the phases of lasers $i$ and $j$ mapped to orientation angle of spins $i$ and $j$, and $\langle . \rangle $ denotes averaging over the ensemble of simulators, that is achieved simultaneously with our coupled lasers.

\begin{figure}[htbp]
\includegraphics[width=85mm]{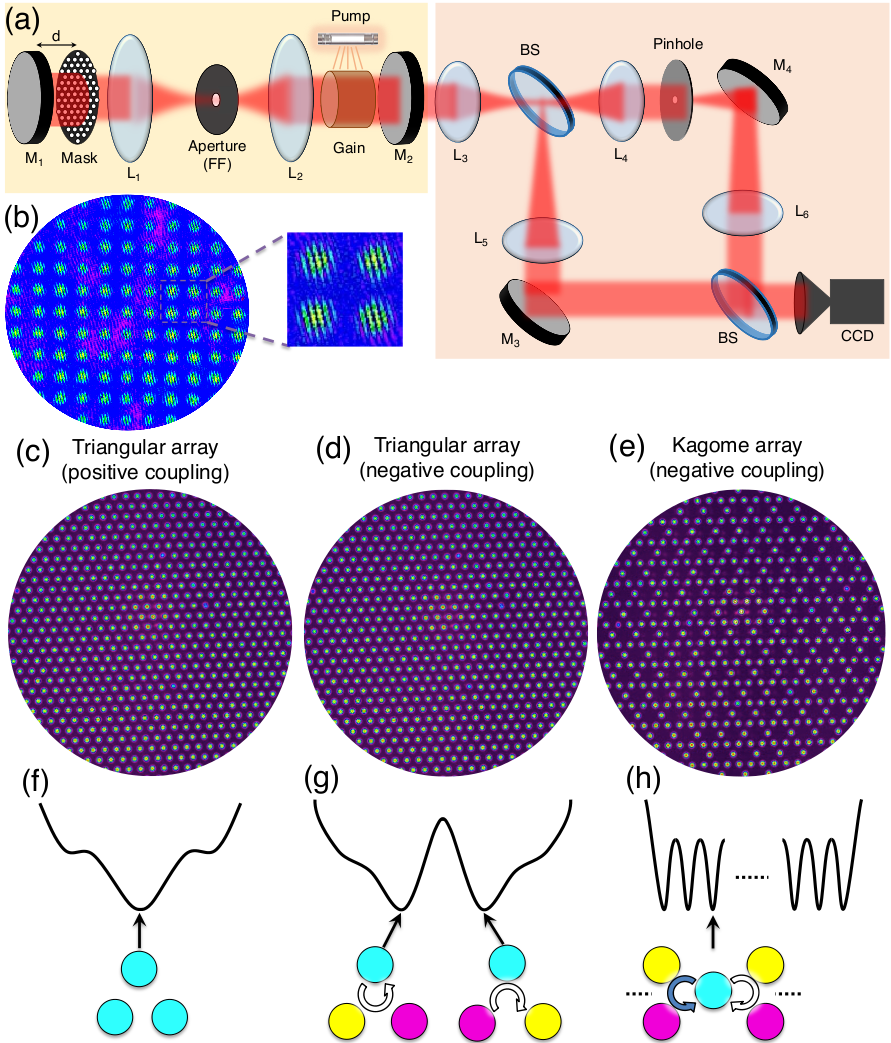}
\caption{\label{fig1}Experimental arrangement, lattice geometries, and representative results. (a) Schematic of a degenerate cavity laser (shaded in yellow) that forms and phase locks lasers in different lattice geometries, together with Mach-Zehnder interferometer (shaded in orange) for analyzing the coherence between the lasers. (b) Experimentally measured interference pattern when a single reference laser interferes with itself and with all the other lasers in the square lattice. Experimental near-field intensity patterns for (c) triangular lattice (positive coupling); (d) triangular lattice (negative coupling); and (e) Kagome lattice (negative coupling). (f) Landscape with a single ground state, corresponding to in-phase locked triangular lattice. (g) Landscape with two degenerate ground states, corresponding to vortex and anti-vortex states of out-of-phase locked triangular lattice. (g) Landscape with highly degenerate ground states, corresponding to $2^{n}$ states ($n$, the number of triangles) in the out-of-phase locked Kagome lattice. Note, for $n=2$, only $1$ state out of $4$ states is shown. Different colors of the lasers denote different values of the phases. Cyan $=0$; yellow $=2\pi/3$; and pink $=-2\pi/3$. M$_{1}$ and M$_{2}$ denote high reflectivity and partial reflectivity cavity mirrors, M$_{3}$ and M$_{4}$ high reflectivity mirrors. L$_{1}$, L$_{2}$, L$_{3}$, L$_{4}$, L$_{5}$, L$_{6}$ plano-convex lenses, BS beam splitter, and CCD camera.}
\end{figure}

The experimental arrangement, array configurations and representative results are presented in Fig.\,\ref{fig1}. Our coupled lasers in lattices are formed in a degenerate cavity shown schematically in Fig.\,\ref{fig1}(a) (yellow shaded region). It is comprised of two mirrors, two lenses in a 4f telescope, a mask containing several hundred circular holes in different lattice geometries, and a Nd:YAG gain medium pumped by a 100 $\mu$s pulsed Xenon flash lamp. The intra-cavity 4f telescope ensures that any field distribution at the mask plane is imaged onto itself after every round-trip. Accordingly, each hole on the mask defines an independent individual laser \cite{Nixon13, Vishwa17, Vishwa15}. Each laser lases with a nearly pure single Gaussian transverse mode (forced by a 200 $\mu$m diameter circular aperture located in the Fourier plane of the intra-cavity telescope) and approximately 250 longitudinal modes that are common to all lasers due to the degenerate cavity condition \cite{Ronen2018}. 
 We verified that lasers are independent, by showing that each laser is incoherent with all the other lasers \cite{Vishwa17,Supplementary}. Coupling between adjacent lasers is introduced by shifting mirror M$_{1}$ a distance $d$ of quarter-Talbot length away from the mask  \cite{Mehuys91}. Such a distance results in negative coupling between adjacent lasers, corresponding to anti-ferromagnetic ordering of classical XY spins \cite{Chene15}. Alternatively, $d$ of half-Talbot length combined with Fourier filtering provides positive coupling between adjacent lasers, corresponding to ferromagnetic ordering \cite{Chene17}.
 
The coherence between the lasers (Eq.\,(1)) is measured using a Mach-Zehnder interferometer, shown in Fig.\,\ref{fig1}(a) (orange shaded region). The output of lasers from the degenerate cavity splits into two channels at the first beam splitter. In one channel, the output of all the lasers is imaged directly onto the CCD camera. In the other channel, a single reference laser is selected using a pinhole of size 50 $\mu$m, and then its light is expanded so that it fully overlaps and interferes with the light of all the lasers with a second beam splitter on the camera. Thus, a single selected reference laser interferes with itself and with all the other lasers. A small tilt between the two channels provides few interference fringes for each laser (exemplified in Fig.\,\ref{fig1}(b) for a square lattice) from which the fringe visibility (coherence) and shift (phase difference) are obtained for all lasers by digital Fourier analysis. The measured coherence function is normalized such that the coherence of the reference laser with itself is one. We also measure the far-field diffraction pattern of the lasers in lattice that corresponds to the ensemble averaged structure factor of the lattice \cite{Nixon13} where sharp Bragg peaks indicate long range phase ordering between the lasers.

\begin{figure}[htbp]
\centering
\includegraphics[width=84mm]{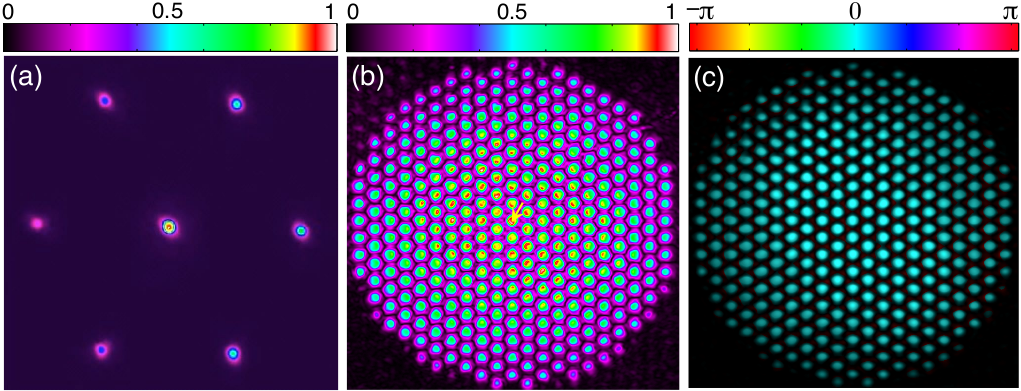}
\caption{Positively coupled lasers in a triangular lattice. (a) Ensemble averaged far-field diffraction pattern, indicating long range in-phase ordering. (b)The coherence calculated from the interference pattern measured by Mach-Zehnder interferometer \cite{Supplementary}, also indicating long range phase ordering. (c) The phases of lasers calculated from the measured interference pattern, indicating long range in-phase ordering throughout the lattice. The yellow arrow in Fig.\,\ref{fig2}(b) denotes the location of reference laser (same in other figures).}
\label{fig2} 
\end{figure}

Using the experimental arrangement shown in Fig.\,\ref{fig1}, we performed a series of experiments to demonstrate fair sampling of ground state manifold in square, triangular and Kagome lattices. We first phase locked about $320$ lasers with positive coupling in a triangular lattice (Fig.\,\ref{fig1}(c)). The results in Fig.\,\ref{fig2} represent an ensemble averaging over about $250$ independent realizations, each corresponding to a different longitudinal mode.
Figure\,\ref{fig2}(a) shows the far-field diffraction pattern of the lasers, where the sharp Bragg peaks indicate long range in-phase ordering. The measured coherence (Fig.\,\ref{fig2}(b)) of the lasers also evidences long range phase ordering and barely decays with distance from the reference laser. Finally, Fig.\,\ref{fig2}(c) shows the measured phases of the lasers (relative to the reference laser), confirming in-phase ordering throughout the lattice. The observed long range in-phase ordering evidences that the entire ensemble of experiments (realized by multiple longitudinal modes) converged to the same non-degenerate ground state, as expected from its single minimal loss manifold illustrated in Fig.\,\ref{fig1}(f). This convergence is analogous to perfect ferromagnetic spin ordering of $XY$ spins. We obtained long range (out-of-phase) ordering also for a square lattice with negative coupling (Fig.\,\ref{fig1}(b)) that has the same single minimal loss manifold \cite{Supplementary}.

\begin{figure}[htbp]
\centering
\includegraphics[width=84mm]{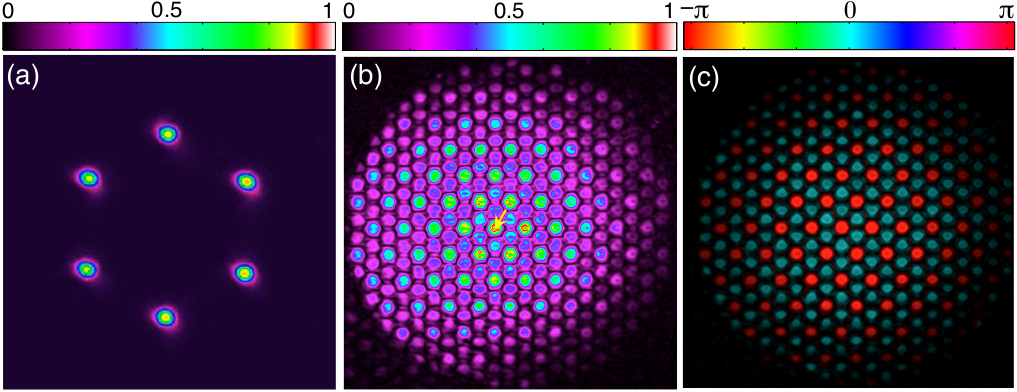}
\caption{Negatively coupled lasers in a triangular lattice. (a) Ensemble averaged far-field diffraction pattern containing two states (vortex and anti-vortex). (b) The coherence calculated from the measured interference pattern \cite{Supplementary}, showing high coherence every three lasers in the ensemble average. (c) The phases of lasers calculated from the measured interference pattern, showing phase difference of $\pi$ (from the reference laser) between nearest neighbors in the ensemble average.}
\label{fig3} 
\end{figure}

Next, we investigated the triangular lattice of about 320  negative coupled lasers. Figure\,\ref{fig3}(a) shows the far-field diffraction pattern, which is comprised of six sharp Bragg peaks that indicate long range phase ordering. Three of these Bragg peaks correspond to a vortex state illustrated as the left ground state in Fig.\,\ref{fig1}(g) and other three peaks correspond to an anti-vortex state illustrated as the right ground state in Fig.\,\ref{fig1}(g) \cite{Nixon13, Chene17,Acomment}. These two degenerate ground states have (by symmetry) an equal probability, to be populated by the ensemble of experiments realized by the multiple longitudinal modes, as indicated by the equal intensity of their Bragg peaks in the diffraction pattern of Fig.\,\ref{fig3}(a).

The co-existence of the two degenerate ground states has remarkable consequences on the measured coherence function that oscillates where the coherence with respect to the reference laser revives every three lasers (Fig.\,\ref{fig3}(b)). This surprising behavior can be understood by noting that for the nearest neighbor (NN) and the next-nearest neighbor (NNN) lasers, the vortex and anti-vortex states differ by $\pm ~2\pi/3$. So their interference fringes are shifted and as result of ensemble averaging reduce the coherence to $50\%$. However, for the next-next-nearest neighbor (NNNN) laser these two states have the same relative phase, yielding coherence of $100\%$ (and then the same $50\%$, $50\%$, $100\%$ coherence periodicity repeated). Figure\,\ref{fig3}(c) shows the ensemble-averaged phases of the lasers, indicating phase difference of $\pi$ (from the reference laser) between the nearest neighbors. This is analogous to XY spin systems, where XY spins oriented at $+~2\pi/3$ and $-~2\pi/3$ yield a magnitude half spin oriented at $\pi$ in the ensemble averaging. 

\begin{figure}[htbp]
\centering
\includegraphics[width=84mm]{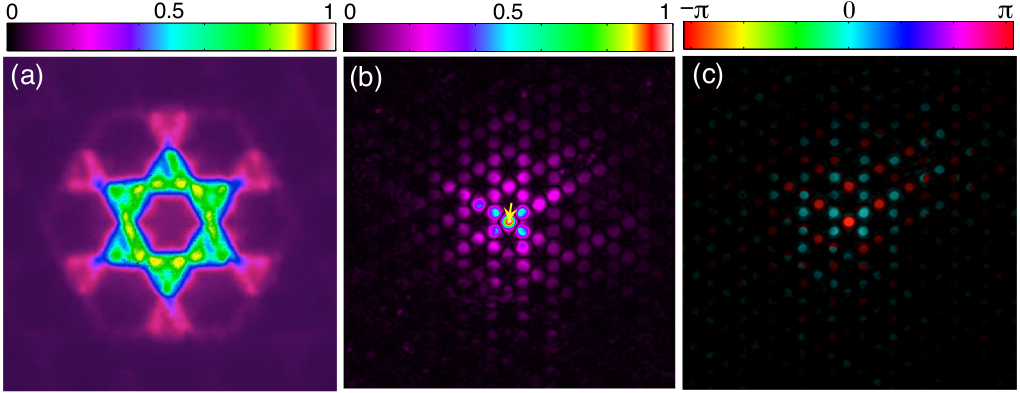}
\caption{Negatively coupled lasers in a Kagome lattice. (a) Ensemble averaged far-field diffraction pattern, containing many vortex and anti-vortex states. (b) The coherence calculated from the measured interference pattern \cite{Supplementary}, showing exponential decaying behavior in the ensemble average. (c) The phases of the lasers calculated from the measured interference pattern, showing phase difference of $\pi$ (from the reference laser) between nearest neighbors in the ensemble average.}
\label{fig4} 
\end{figure}

The XY spin Hamiltonian on a Kagome lattice (Fig.\,\ref{fig1}(e)) exhibits highly non-trivial features such as geometric frustration that arises due to massive degeneracy in it's ground state \cite{Nixon13, Chalker92, Chalker98}. The degeneracy scales exponentially with the system size, thus performing fair sampling is a computationally hard problem \cite{Tamate16, Takeda18}. The results for fair sampling in a Kagome lattice with about $350$ negatively coupled lasers are shown in Fig.\,\ref{fig4}. 

Figure\,\ref{fig4}(a) shows the ensemble averaged far-field diffraction pattern that consists of large area Bragg lobes (rather than sharp peaks) indicating lack of long range phase ordering, in agreement with the theoretical results \cite{Nixon13, Chalker98}. Specifically, in Kagome lattice, each triangle of lasers can randomly find either the vortex or the anti-vortex degenerate ground state with equal probability, thereby suppressing long range phase ordering. Figure\,\ref{fig4}(b) shows a rapid decay of coherence, indicating again lack of long range phase ordering. The coherence function can be quantitatively determined by calculating the probability distribution of states that have relative phase differences $\pm 2\pi/3$ or $0$ with respect to the reference laser. For example, for NN, equal probability states with relative phases $\pm 2\pi/3$, reduce the coherence to $50\%$. For NNN and NNNN, the coherence is reduced to $25\%$ and $12.5\%$, respectively \cite{Bcomment}. More generally, at a distance $n$ from the reference laser, the coherence continues to drop exponentially as $1/2^{n}$. Figure\,\ref{fig4}(c) shows the ensemble averaged phase ordering, indicating a $\pi$ (from the reference laser) phase difference between the nearest neighbors, analogous to the XY spin system.

\begin{figure}[htbp]
\centering
\includegraphics[width=85mm]{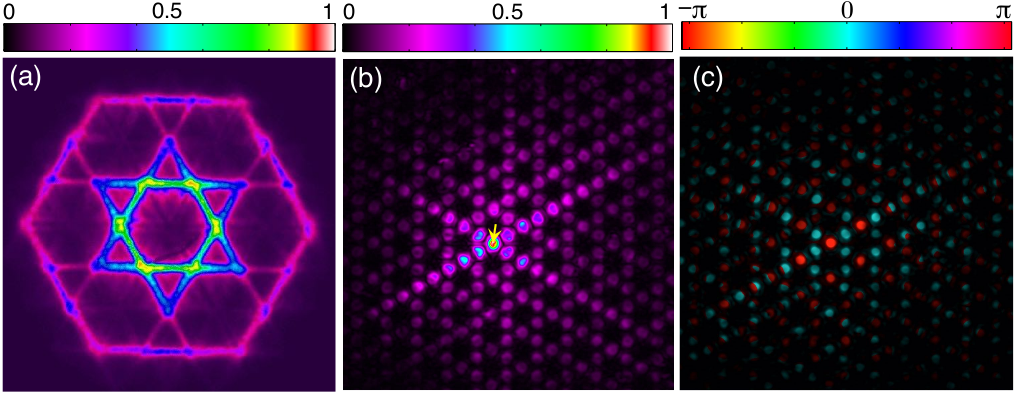}
\caption{Negatively coupled lasers in a Kagome lattice with a large diameter intracavity Fourier aperture. (a) Ensemble averaged far-field diffraction pattern. (b) The coherence calculated from the measured interference pattern \cite{Supplementary}. (c) The phases of the lasers calculated from the measured interference pattern.}
\label{fig5} 
\end{figure}

We also investigated the coherence of the Kagome lattice when the intra-cavity Fourier aperture has large diameter, so that each laser is no longer a pure $\mathrm{TEM}_{00}$ mode, and contains fine internal features which diffracts faster and can generate NNN coupling. The results, shown in Fig.\,\ref{fig5} differ dramatically from those without NNN coupling of Fig.\,\ref{fig4}. Fig.\,\ref{fig5}(a) presents the ensemble averaged far-field diffraction pattern, which now consists of sharp narrow lines indicating long range phase ordering only along certain directions. Figure\,\ref{fig5}(b) shows the measured coherence that decays slowly along certain directions, confirming such anisotropic long range phase ordering. Figure \ref{fig5}(c) shows the ensemble averaged phases, which again indicates the relative phase difference of $\pi$ (from the reference laser) between nearest neighbors. This intriguing, highly directional phase ordering is also observed in our numerical simulations \cite{Supplementary} and is accompanied by spontaneous intensity pattern formation as in the stripe phase of ultracold atoms \cite{Keterle17}. 
 
\begin{figure}[htbp]
\centering
\includegraphics[width=86mm]{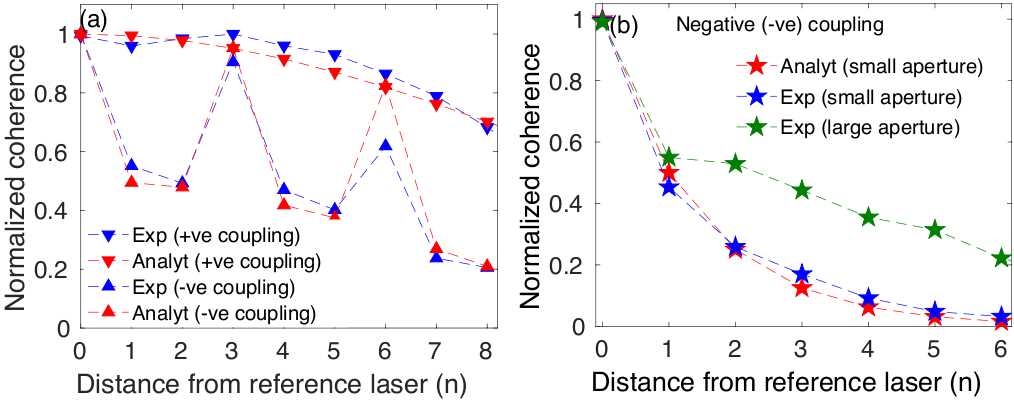}
\caption{The ensemble averaged normalized coherence as a function of distance from the reference laser. (a) Normalized coherence for the positively coupled lasers (inverted blue and red triangles) and negatively coupled lasers (blue and red triangles) in a triangular lattice. For the positive coupling, the coherence is monotonic, whereas for the negative coupling the coherence shows an oscillatory behavior in agreement with the analytical results. (b) In a negatively coupled Kagome lattice, normalized coherence decays exponentially for NN coupling (blue and red stars) but decay much slower with NNN coupling (green stars).}
\label{fig6} 
\end{figure}

Finally, we have quantified the coherence as a function of distance from the reference laser, and compared the experimental results to the analytical ones \cite{Supplementary}, as shown in Fig.\,\ref{fig6}.
Figure\,\ref{fig6}(a) shows the normalized coherence as a function of the distance from the reference laser for positively and negatively coupled lasers in a triangular lattice. For the positively coupled lasers, the coherence decays slowly and monotonically (blue and red inverted triangles). As evident, the ensemble averaging does not reduce the coherence indicating a single non-degenerate ground state. For the negatively coupled lasers, both the analytical and experimental coherence show an oscillatory behavior as a function of distance from the reference laser (blue and red triangles), where the coherence revives every three lasers. This loss and revival of the ensemble averaged coherence with distance from the reference laser indicates two degenerate ground states. 

Figure\,\ref{fig6}(b) shows the normalized coherence as a function of the distance from the reference laser in a Kagome lattice with NN and NNN negative coupling. For NN negative coupling, the ensemble averaged coherence decays exponentially, in agreement with the analytical results \cite{Bcomment} (blue and red stars). The exponential decay indicates massive degeneracy in the ground state that scales exponentially with the system size due to geometric frustration \cite{Nixon13}. For NNN coupling, the ensemble averaged coherence decays much slower as a function of the distance from the reference laser (green stars), indicating reduced number of degenerate ground states.

In conclusion, we presented a simulator based on dissipatively coupled lasers for rapid and efficient fair sampling of magnetic ordering of XY spin Hamiltonian with ground state degeneracy. The simulator exploited 250 longitudinal modes of each laser to form an ensemble of 250 identical but independent simulators so as to provide massive parallelism in performing statistical fair sampling. We investigated the ground state manifold in different geometries such as square, triangular and Kagome lattices. For negative (positive) coupling, we observed a single ground state for the square (triangular) lattice, two degenerate ground states for the triangular lattice, and geometrically frustrated highly degenerate ground states for the Kagome lattice. For these cases, the corresponding spatial coherence functions are near-uniform, oscillatory and exponentially decaying. Under certain conditions, we also observed highly directional phase ordering in a Kagome lattice, indicating reduced ground state degeneracy. Our simulator with rapid fair sampling of ground state manifold could potentially be exploited to address various combinatorial optimization problems. We plan to extend our work to study the effects of defects on the ensemble averaged coherence in 2D lattices, and their influence on the ground state manifolds.

\begin{acknowledgments}
The authors acknowledge partial support from  Israel Science Foundation (ISF) and ISIRD grant from Indian Institute of Technology Ropar.
\end{acknowledgments}

\nocite{*}

\begin{thebibliography}{0}%
\makeatletter
\providecommand \@ifxundefined [1]{%
 \@ifx{#1\undefined}
}%
\providecommand \@ifnum [1]{%
 \ifnum #1\expandafter \@firstoftwo
 \else \expandafter \@secondoftwo
 \fi
}%
\providecommand \@ifx [1]{%
 \ifx #1\expandafter \@firstoftwo
 \else \expandafter \@secondoftwo
 \fi
}%
\providecommand \natexlab [1]{#1}%
\providecommand \enquote  [1]{``#1''}%
\providecommand \bibnamefont  [1]{#1}%
\providecommand \bibfnamefont [1]{#1}%
\providecommand \citenamefont [1]{#1}%
\providecommand \href@noop [0]{\@secondoftwo}%
\providecommand \href [0]{\begingroup \@sanitize@url \@href}%
\providecommand \@href[1]{\@@startlink{#1}\@@href}%
\providecommand \@@href[1]{\endgroup#1\@@endlink}%
\providecommand \@sanitize@url [0]{\catcode `\\12\catcode `\$12\catcode
  `\&12\catcode `\#12\catcode `\^12\catcode `\_12\catcode `\%12\relax}%
\providecommand \@@startlink[1]{}%
\providecommand \@@endlink[0]{}%
\providecommand \url  [0]{\begingroup\@sanitize@url \@url }%
\providecommand \@url [1]{\endgroup\@href {#1}{\urlprefix }}%
\providecommand \urlprefix  [0]{URL }%
\providecommand \Eprint [0]{\href }%
\providecommand \doibase [0]{https://doi.org/}%
\providecommand \selectlanguage [0]{\@gobble}%
\providecommand \bibinfo  [0]{\@secondoftwo}%
\providecommand \bibfield  [0]{\@secondoftwo}%
\providecommand \translation [1]{[#1]}%
\providecommand \BibitemOpen [0]{}%
\providecommand \bibitemStop [0]{}%
\providecommand \bibitemNoStop [0]{.\EOS\space}%
\providecommand \EOS [0]{\spacefactor3000\relax}%
\providecommand \BibitemShut  [1]{\csname bibitem#1\endcsname}%
\let\auto@bib@innerbib\@empty
\end{thebibliography}%


\begin{thebibliography}{34}%
\makeatletter
\providecommand \@ifxundefined [1]{%
 \@ifx{#1\undefined}
}%
\providecommand \@ifnum [1]{%
 \ifnum #1\expandafter \@firstoftwo
 \else \expandafter \@secondoftwo
 \fi
}%
\providecommand \@ifx [1]{%
 \ifx #1\expandafter \@firstoftwo
 \else \expandafter \@secondoftwo
 \fi
}%
\providecommand \natexlab [1]{#1}%
\providecommand \enquote  [1]{``#1''}%
\providecommand \bibnamefont  [1]{#1}%
\providecommand \bibfnamefont [1]{#1}%
\providecommand \citenamefont [1]{#1}%
\providecommand \href@noop [0]{\@secondoftwo}%
\providecommand \href [0]{\begingroup \@sanitize@url \@href}%
\providecommand \@href[1]{\@@startlink{#1}\@@href}%
\providecommand \@@href[1]{\endgroup#1\@@endlink}%
\providecommand \@sanitize@url [0]{\catcode `\\12\catcode `\$12\catcode
  `\&12\catcode `\#12\catcode `\^12\catcode `\_12\catcode `\%12\relax}%
\providecommand \@@startlink[1]{}%
\providecommand \@@endlink[0]{}%
\providecommand \url  [0]{\begingroup\@sanitize@url \@url }%
\providecommand \@url [1]{\endgroup\@href {#1}{\urlprefix }}%
\providecommand \urlprefix  [0]{URL }%
\providecommand \Eprint [0]{\href }%
\providecommand \doibase [0]{https://doi.org/}%
\providecommand \selectlanguage [0]{\@gobble}%
\providecommand \bibinfo  [0]{\@secondoftwo}%
\providecommand \bibfield  [0]{\@secondoftwo}%
\providecommand \translation [1]{[#1]}%
\providecommand \BibitemOpen [0]{}%
\providecommand \bibitemStop [0]{}%
\providecommand \bibitemNoStop [0]{.\EOS\space}%
\providecommand \EOS [0]{\spacefactor3000\relax}%
\providecommand \BibitemShut  [1]{\csname bibitem#1\endcsname}%
\let\auto@bib@innerbib\@empty
\bibitem [{\citenamefont {Papadimitriou}\ and\ \citenamefont
  {Steiglits}(1998)}]{Steiglits98}%
  \BibitemOpen
  \bibfield  {author} {\bibinfo {author} {\bibfnamefont {C.~H.}\ \bibnamefont
  {Papadimitriou}}\ and\ \bibinfo {author} {\bibfnamefont {K.}~\bibnamefont
  {Steiglits}},\ }\href@noop {} {\emph {\bibinfo {title} {Combinatorial
  Optimization: Algorithims and Complexity}}}\ (\bibinfo  {publisher} {Dover
  Publications},\ \bibinfo {year} {1998})\BibitemShut {NoStop}%
\bibitem [{\citenamefont {Yamamoto}\ \emph {et~al.}(2017)\citenamefont
  {Yamamoto}, \citenamefont {Aihara}, \citenamefont {Leleu}, \citenamefont
  {Kawarabayashi}, \citenamefont {Kako}, \citenamefont {Fejer}, \citenamefont
  {Inoue},\ and\ \citenamefont {Takesue}}]{Yamamoto17}%
  \BibitemOpen
  \bibfield  {author} {\bibinfo {author} {\bibfnamefont {Y.}~\bibnamefont
  {Yamamoto}}, \bibinfo {author} {\bibfnamefont {K.}~\bibnamefont {Aihara}},
  \bibinfo {author} {\bibfnamefont {T.}~\bibnamefont {Leleu}}, \bibinfo
  {author} {\bibfnamefont {K.}~\bibnamefont {Kawarabayashi}}, \bibinfo {author}
  {\bibfnamefont {S.}~\bibnamefont {Kako}}, \bibinfo {author} {\bibfnamefont
  {M.}~\bibnamefont {Fejer}}, \bibinfo {author} {\bibfnamefont
  {K.}~\bibnamefont {Inoue}},\ and\ \bibinfo {author} {\bibfnamefont
  {H.}~\bibnamefont {Takesue}},\ }\bibfield  {title} {\bibinfo {title}
  {Coherent ising machines—optical neural networks operating at the quantum
  limit},\ }\href@noop {} {\bibfield  {journal} {\bibinfo  {journal} {npj\
  Quantum\ Information}\ }\textbf {\bibinfo {volume} {3}},\ \bibinfo {pages}
  {49} (\bibinfo {year} {2017})}\BibitemShut {NoStop}%
\bibitem [{\citenamefont {Berloff}\ \emph {et~al.}(2017)\citenamefont
  {Berloff}, \citenamefont {Silva}, \citenamefont {Kalinin}, \citenamefont
  {Askitopoulos}, \citenamefont {Topfer}, \citenamefont {Cilibrizzi},
  \citenamefont {Langbein},\ and\ \citenamefont {Lagoudakis}}]{Berloff17}%
  \BibitemOpen
  \bibfield  {author} {\bibinfo {author} {\bibfnamefont {N.~G.}\ \bibnamefont
  {Berloff}}, \bibinfo {author} {\bibfnamefont {M.}~\bibnamefont {Silva}},
  \bibinfo {author} {\bibfnamefont {K.}~\bibnamefont {Kalinin}}, \bibinfo
  {author} {\bibfnamefont {A.}~\bibnamefont {Askitopoulos}}, \bibinfo {author}
  {\bibfnamefont {J.~D.}\ \bibnamefont {Topfer}}, \bibinfo {author}
  {\bibfnamefont {P.}~\bibnamefont {Cilibrizzi}}, \bibinfo {author}
  {\bibfnamefont {W.}~\bibnamefont {Langbein}},\ and\ \bibinfo {author}
  {\bibfnamefont {P.~G.}\ \bibnamefont {Lagoudakis}},\ }\bibfield  {title}
  {\bibinfo {title} {Realizing the classical xy hamiltonian in polariton
  simulators},\ }\href@noop {} {\bibfield  {journal} {\bibinfo  {journal}
  {Nat.\ Mater.}\ }\textbf {\bibinfo {volume} {16}},\ \bibinfo {pages} {1120}
  (\bibinfo {year} {2017})}\BibitemShut {NoStop}%
\bibitem [{\citenamefont {Lagoudakis}\ and\ \citenamefont
  {Berloff}(2017)}]{Lagoudakis17}%
  \BibitemOpen
  \bibfield  {author} {\bibinfo {author} {\bibfnamefont {P.~G.}\ \bibnamefont
  {Lagoudakis}}\ and\ \bibinfo {author} {\bibfnamefont {N.~G.}\ \bibnamefont
  {Berloff}},\ }\bibfield  {title} {\bibinfo {title} {A polariton graph
  simulator},\ }\href@noop {} {\bibfield  {journal} {\bibinfo  {journal} {New.\
  J.\ Phys.}\ }\textbf {\bibinfo {volume} {19}},\ \bibinfo {pages} {125008}
  (\bibinfo {year} {2017})}\BibitemShut {NoStop}%
\bibitem [{\citenamefont {McMahon}\ \emph {et~al.}(2016)\citenamefont
  {McMahon}, \citenamefont {Marandi}, \citenamefont {Haribara}, \citenamefont
  {Hamerly}, \citenamefont {Lagrock}, \citenamefont {Tamate}, \citenamefont
  {Inagaki}, \citenamefont {Takesue}, \citenamefont {Utsunomiya}, \citenamefont
  {Aihara}, \citenamefont {Byer}, \citenamefont {Fejer}, \citenamefont
  {Mabuchi},\ and\ \citenamefont {Yamamoto}}]{McMahon16}%
  \BibitemOpen
  \bibfield  {author} {\bibinfo {author} {\bibfnamefont {P.~L.}\ \bibnamefont
  {McMahon}}, \bibinfo {author} {\bibfnamefont {A.}~\bibnamefont {Marandi}},
  \bibinfo {author} {\bibfnamefont {Y.}~\bibnamefont {Haribara}}, \bibinfo
  {author} {\bibfnamefont {R.}~\bibnamefont {Hamerly}}, \bibinfo {author}
  {\bibfnamefont {C.}~\bibnamefont {Lagrock}}, \bibinfo {author} {\bibfnamefont
  {S.}~\bibnamefont {Tamate}}, \bibinfo {author} {\bibfnamefont
  {T.}~\bibnamefont {Inagaki}}, \bibinfo {author} {\bibfnamefont
  {H.}~\bibnamefont {Takesue}}, \bibinfo {author} {\bibfnamefont
  {S.}~\bibnamefont {Utsunomiya}}, \bibinfo {author} {\bibfnamefont
  {K.}~\bibnamefont {Aihara}}, \bibinfo {author} {\bibfnamefont
  {R.}~\bibnamefont {Byer}}, \bibinfo {author} {\bibfnamefont {M.~M.}\
  \bibnamefont {Fejer}}, \bibinfo {author} {\bibfnamefont {H.}~\bibnamefont
  {Mabuchi}},\ and\ \bibinfo {author} {\bibfnamefont {Y.}~\bibnamefont
  {Yamamoto}},\ }\bibfield  {title} {\bibinfo {title} {A fully programmable
  100-spin coherent ising machine with all-to-all connections},\ }\href@noop {}
  {\bibfield  {journal} {\bibinfo  {journal} {Science}\ }\textbf {\bibinfo
  {volume} {354}},\ \bibinfo {pages} {614} (\bibinfo {year}
  {2016})}\BibitemShut {NoStop}%
\bibitem [{\citenamefont {Takeda}\ \emph {et~al.}(2018)\citenamefont {Takeda},
  \citenamefont {Tamate}, \citenamefont {Yamamoto}, \citenamefont {Takesue},
  \citenamefont {Inagaki},\ and\ \citenamefont {Utsunomiya}}]{Takeda18}%
  \BibitemOpen
  \bibfield  {author} {\bibinfo {author} {\bibfnamefont {Y.}~\bibnamefont
  {Takeda}}, \bibinfo {author} {\bibfnamefont {S.}~\bibnamefont {Tamate}},
  \bibinfo {author} {\bibfnamefont {Y.}~\bibnamefont {Yamamoto}}, \bibinfo
  {author} {\bibfnamefont {H.}~\bibnamefont {Takesue}}, \bibinfo {author}
  {\bibfnamefont {T.}~\bibnamefont {Inagaki}},\ and\ \bibinfo {author}
  {\bibfnamefont {S.}~\bibnamefont {Utsunomiya}},\ }\bibfield  {title}
  {\bibinfo {title} {Boltzmann sampling for an xy model using a non-degenerate
  optical parametric oscillator network},\ }\href@noop {} {\bibfield  {journal}
  {\bibinfo  {journal} {Quantum\ Sci.\ Technol.}\ }\textbf {\bibinfo {volume}
  {3}},\ \bibinfo {pages} {014004} (\bibinfo {year} {2018})}\BibitemShut
  {NoStop}%
\bibitem [{\citenamefont {Nixon}\ \emph {et~al.}(2013)\citenamefont {Nixon},
  \citenamefont {Ronen}, \citenamefont {Friesem},\ and\ \citenamefont
  {Davidson}}]{Nixon13}%
  \BibitemOpen
  \bibfield  {author} {\bibinfo {author} {\bibfnamefont {M.}~\bibnamefont
  {Nixon}}, \bibinfo {author} {\bibfnamefont {E.}~\bibnamefont {Ronen}},
  \bibinfo {author} {\bibfnamefont {A.~A.}\ \bibnamefont {Friesem}},\ and\
  \bibinfo {author} {\bibfnamefont {N.}~\bibnamefont {Davidson}},\ }\bibfield
  {title} {\bibinfo {title} {Observing geometric frustration with thousands of
  coupled lasers},\ }\href@noop {} {\bibfield  {journal} {\bibinfo  {journal}
  {Phys.\ Rev.\ Lett.}\ }\textbf {\bibinfo {volume} {110}},\ \bibinfo {pages}
  {184102} (\bibinfo {year} {2013})}\BibitemShut {NoStop}%
\bibitem [{\citenamefont {Pal}\ \emph {et~al.}(2017)\citenamefont {Pal},
  \citenamefont {Tradonsky}, \citenamefont {Chriki}, \citenamefont {Friesem},\
  and\ \citenamefont {Davidson}}]{Vishwa17}%
  \BibitemOpen
  \bibfield  {author} {\bibinfo {author} {\bibfnamefont {V.}~\bibnamefont
  {Pal}}, \bibinfo {author} {\bibfnamefont {C.}~\bibnamefont {Tradonsky}},
  \bibinfo {author} {\bibfnamefont {R.}~\bibnamefont {Chriki}}, \bibinfo
  {author} {\bibfnamefont {A.~A.}\ \bibnamefont {Friesem}},\ and\ \bibinfo
  {author} {\bibfnamefont {N.}~\bibnamefont {Davidson}},\ }\bibfield  {title}
  {\bibinfo {title} {Observing dissipative topological defects with coupled
  lasers},\ }\href@noop {} {\bibfield  {journal} {\bibinfo  {journal} {Phys.\
  Rev.\ Lett.}\ }\textbf {\bibinfo {volume} {119}},\ \bibinfo {pages} {013902}
  (\bibinfo {year} {2017})}\BibitemShut {NoStop}%
\bibitem [{\citenamefont {Inagaki}\ \emph {et~al.}(2016)\citenamefont
  {Inagaki}, \citenamefont {Inaba}, \citenamefont {Hamerly}, \citenamefont
  {Inoue}, \citenamefont {Yamamoto},\ and\ \citenamefont
  {Takesue}}]{Inagaki16}%
  \BibitemOpen
  \bibfield  {author} {\bibinfo {author} {\bibfnamefont {T.}~\bibnamefont
  {Inagaki}}, \bibinfo {author} {\bibfnamefont {K.}~\bibnamefont {Inaba}},
  \bibinfo {author} {\bibfnamefont {R.}~\bibnamefont {Hamerly}}, \bibinfo
  {author} {\bibfnamefont {K.}~\bibnamefont {Inoue}}, \bibinfo {author}
  {\bibfnamefont {Y.}~\bibnamefont {Yamamoto}},\ and\ \bibinfo {author}
  {\bibfnamefont {H.}~\bibnamefont {Takesue}},\ }\bibfield  {title} {\bibinfo
  {title} {Large-scale ising spin network based on degenerate optical
  parametric oscillators},\ }\href@noop {} {\bibfield  {journal} {\bibinfo
  {journal} {Nat.\ Photon.}\ }\textbf {\bibinfo {volume} {10}},\ \bibinfo
  {pages} {415} (\bibinfo {year} {2016})}\BibitemShut {NoStop}%
\bibitem [{\citenamefont {Marandi}\ \emph {et~al.}(2014)\citenamefont
  {Marandi}, \citenamefont {Wang}, \citenamefont {Takata}, \citenamefont
  {Byer},\ and\ \citenamefont {Yamamoto}}]{Marandi14}%
  \BibitemOpen
  \bibfield  {author} {\bibinfo {author} {\bibfnamefont {A.}~\bibnamefont
  {Marandi}}, \bibinfo {author} {\bibfnamefont {Z.}~\bibnamefont {Wang}},
  \bibinfo {author} {\bibfnamefont {K.}~\bibnamefont {Takata}}, \bibinfo
  {author} {\bibfnamefont {R.~L.}\ \bibnamefont {Byer}},\ and\ \bibinfo
  {author} {\bibfnamefont {Y.}~\bibnamefont {Yamamoto}},\ }\bibfield  {title}
  {\bibinfo {title} {Network of time-multiplexed optical parametric oscillators
  as a coherent ising machine},\ }\href@noop {} {\bibfield  {journal} {\bibinfo
   {journal} {Nat.\ Photon.}\ }\textbf {\bibinfo {volume} {8}},\ \bibinfo
  {pages} {937} (\bibinfo {year} {2014})}\BibitemShut {NoStop}%
\bibitem [{\citenamefont {Tamate}\ \emph {et~al.}(2016)\citenamefont {Tamate},
  \citenamefont {Yamamoto}, \citenamefont {Marandi}, \citenamefont {McMahon},\
  and\ \citenamefont {Utsunomiya}}]{Tamate16}%
  \BibitemOpen
  \bibfield  {author} {\bibinfo {author} {\bibfnamefont {S.}~\bibnamefont
  {Tamate}}, \bibinfo {author} {\bibfnamefont {Y.}~\bibnamefont {Yamamoto}},
  \bibinfo {author} {\bibfnamefont {A.}~\bibnamefont {Marandi}}, \bibinfo
  {author} {\bibfnamefont {P.}~\bibnamefont {McMahon}},\ and\ \bibinfo {author}
  {\bibfnamefont {S.}~\bibnamefont {Utsunomiya}},\ }\bibfield  {title}
  {\bibinfo {title} {Simulating the classical xy model with a laser network},\
  }\href@noop {} {\bibfield  {journal} {\bibinfo  {journal} {arXiv:1608.00358}\
  } (\bibinfo {year} {2016})}\BibitemShut {NoStop}%
\bibitem [{\citenamefont {K$\mathrm{\ddot{o}}$nz}\ \emph
  {et~al.}(2019)\citenamefont {K$\mathrm{\ddot{o}}$nz}, \citenamefont
  {Mazzola}, \citenamefont {Ochoa}, \citenamefont {Katzgraber},\ and\
  \citenamefont {Troyer}}]{Konz19}%
  \BibitemOpen
  \bibfield  {author} {\bibinfo {author} {\bibfnamefont {M.~S.}\ \bibnamefont
  {K$\mathrm{\ddot{o}}$nz}}, \bibinfo {author} {\bibfnamefont {G.}~\bibnamefont
  {Mazzola}}, \bibinfo {author} {\bibfnamefont {A.~J.}\ \bibnamefont {Ochoa}},
  \bibinfo {author} {\bibfnamefont {H.~G.}\ \bibnamefont {Katzgraber}},\ and\
  \bibinfo {author} {\bibfnamefont {M.}~\bibnamefont {Troyer}},\ }\bibfield
  {title} {\bibinfo {title} {Uncertain fate of fair sampling in quantum
  annealing},\ }\href@noop {} {\bibfield  {journal} {\bibinfo  {journal}
  {Phys.\ Rev.\ A}\ }\textbf {\bibinfo {volume} {100}},\ \bibinfo {pages}
  {030303(R)} (\bibinfo {year} {2019})}\BibitemShut {NoStop}%
\bibitem [{\citenamefont {Mandr$\mathrm{\grave{a}}$}\ \emph
  {et~al.}(2017)\citenamefont {Mandr$\mathrm{\grave{a}}$}, \citenamefont
  {Zhu},\ and\ \citenamefont {Katzgraber}}]{Mandra17}%
  \BibitemOpen
  \bibfield  {author} {\bibinfo {author} {\bibfnamefont {S.}~\bibnamefont
  {Mandr$\mathrm{\grave{a}}$}}, \bibinfo {author} {\bibfnamefont
  {Z.}~\bibnamefont {Zhu}},\ and\ \bibinfo {author} {\bibfnamefont {H.~G.}\
  \bibnamefont {Katzgraber}},\ }\bibfield  {title} {\bibinfo {title}
  {Exponentially biased ground-state sampling of quantum annealing machines
  with transverse-field driving hamiltonians},\ }\href@noop {} {\bibfield
  {journal} {\bibinfo  {journal} {Phys.\ Rev.\ Lett.}\ }\textbf {\bibinfo
  {volume} {118}},\ \bibinfo {pages} {070502} (\bibinfo {year}
  {2017})}\BibitemShut {NoStop}%
\bibitem [{\citenamefont {Farhi}\ \emph {et~al.}(2000)\citenamefont {Farhi},
  \citenamefont {Goldstone}, \citenamefont {Gutmann},\ and\ \citenamefont
  {Sipser}}]{Farhi}%
  \BibitemOpen
  \bibfield  {author} {\bibinfo {author} {\bibfnamefont {E.}~\bibnamefont
  {Farhi}}, \bibinfo {author} {\bibfnamefont {J.}~\bibnamefont {Goldstone}},
  \bibinfo {author} {\bibfnamefont {S.}~\bibnamefont {Gutmann}},\ and\ \bibinfo
  {author} {\bibfnamefont {M.}~\bibnamefont {Sipser}},\ }\bibfield  {title}
  {\bibinfo {title} {Quantum computation by adiabatic evolution},\ }\href@noop
  {} {\bibfield  {journal} {\bibinfo  {journal} {arXiv:quant-ph/0001106}\ }
  (\bibinfo {year} {2000})}\BibitemShut {NoStop}%
\bibitem [{\citenamefont {Acebr{\'o}n}\ \emph {et~al.}(2005)\citenamefont
  {Acebr{\'o}n}, \citenamefont {Bonilla}, \citenamefont {Vicente},
  \citenamefont {Ritort},\ and\ \citenamefont {Spigler}}]{Acebron05}%
  \BibitemOpen
  \bibfield  {author} {\bibinfo {author} {\bibfnamefont {J.~A.}\ \bibnamefont
  {Acebr{\'o}n}}, \bibinfo {author} {\bibfnamefont {L.~L.}\ \bibnamefont
  {Bonilla}}, \bibinfo {author} {\bibfnamefont {C.~J.~P.}\ \bibnamefont
  {Vicente}}, \bibinfo {author} {\bibfnamefont {F.}~\bibnamefont {Ritort}},\
  and\ \bibinfo {author} {\bibfnamefont {R.}~\bibnamefont {Spigler}},\
  }\bibfield  {title} {\bibinfo {title} {The kuramoto model: A simple paradigm
  for synchronization phenomena},\ }\href@noop {} {\bibfield  {journal}
  {\bibinfo  {journal} {Rev.\ Mod.\ Phys.}\ }\textbf {\bibinfo {volume} {77}},\
  \bibinfo {pages} {137} (\bibinfo {year} {2005})}\BibitemShut {NoStop}%
\bibitem [{\citenamefont {Campo}\ and\ \citenamefont {Zurek}(2014)}]{Zurek14}%
  \BibitemOpen
  \bibfield  {author} {\bibinfo {author} {\bibfnamefont {A.~D.}\ \bibnamefont
  {Campo}}\ and\ \bibinfo {author} {\bibfnamefont {W.~H.}\ \bibnamefont
  {Zurek}},\ }\bibfield  {title} {\bibinfo {title} {Universality of phase
  transition dynamics: Topological defects from symmetry breaking},\
  }\href@noop {} {\bibfield  {journal} {\bibinfo  {journal} {Int.\ J.\ Mod.\
  Phys.\ A}\ }\textbf {\bibinfo {volume} {29}},\ \bibinfo {pages} {1430018}
  (\bibinfo {year} {2014})}\BibitemShut {NoStop}%
\bibitem [{\citenamefont {Mermin}(1979)}]{Mermin79}%
  \BibitemOpen
  \bibfield  {author} {\bibinfo {author} {\bibfnamefont {N.~D.}\ \bibnamefont
  {Mermin}},\ }\bibfield  {title} {\bibinfo {title} {The topological theory of
  defects in ordered media},\ }\href@noop {} {\bibfield  {journal} {\bibinfo
  {journal} {Rev.\ Mod.\ Phys.}\ }\textbf {\bibinfo {volume} {51}},\ \bibinfo
  {pages} {591} (\bibinfo {year} {1979})}\BibitemShut {NoStop}%
\bibitem [{\citenamefont {Coullet}\ \emph {et~al.}(1989)\citenamefont
  {Coullet}, \citenamefont {Gil},\ and\ \citenamefont {Rocca}}]{Coullet89}%
  \BibitemOpen
  \bibfield  {author} {\bibinfo {author} {\bibfnamefont {P.}~\bibnamefont
  {Coullet}}, \bibinfo {author} {\bibfnamefont {L.}~\bibnamefont {Gil}},\ and\
  \bibinfo {author} {\bibfnamefont {F.}~\bibnamefont {Rocca}},\ }\bibfield
  {title} {\bibinfo {title} {Optical vortices},\ }\href@noop {} {\bibfield
  {journal} {\bibinfo  {journal} {Opt.\ Commun.}\ }\textbf {\bibinfo {volume}
  {73}},\ \bibinfo {pages} {403} (\bibinfo {year} {1989})}\BibitemShut
  {NoStop}%
\bibitem [{\citenamefont {Kibble}(1976)}]{Kibble76}%
  \BibitemOpen
  \bibfield  {author} {\bibinfo {author} {\bibfnamefont {T.~W.~B.}\
  \bibnamefont {Kibble}},\ }\bibfield  {title} {\bibinfo {title} {Topology of
  cosmic domains and strings},\ }\href@noop {} {\bibfield  {journal} {\bibinfo
  {journal} {J.\ Phys.\ A: Math.\ Gen.}\ }\textbf {\bibinfo {volume} {9}},\
  \bibinfo {pages} {1387} (\bibinfo {year} {1976})}\BibitemShut {NoStop}%
\bibitem [{\citenamefont {Zurek}(1985)}]{Zurek85}%
  \BibitemOpen
  \bibfield  {author} {\bibinfo {author} {\bibfnamefont {W.~H.}\ \bibnamefont
  {Zurek}},\ }\bibfield  {title} {\bibinfo {title} {Cosmological experiments in
  superfluid helium?},\ }\href@noop {} {\bibfield  {journal} {\bibinfo
  {journal} {Nature}\ }\textbf {\bibinfo {volume} {317}},\ \bibinfo {pages}
  {505} (\bibinfo {year} {1985})}\BibitemShut {NoStop}%
\bibitem [{\citenamefont {Navon}\ \emph {et~al.}(2015)\citenamefont {Navon},
  \citenamefont {Gaunt}, \citenamefont {Smith},\ and\ \citenamefont
  {Hadzibabic}}]{Navon15}%
  \BibitemOpen
  \bibfield  {author} {\bibinfo {author} {\bibfnamefont {N.}~\bibnamefont
  {Navon}}, \bibinfo {author} {\bibfnamefont {A.~L.}\ \bibnamefont {Gaunt}},
  \bibinfo {author} {\bibfnamefont {R.~P.}\ \bibnamefont {Smith}},\ and\
  \bibinfo {author} {\bibfnamefont {Z.}~\bibnamefont {Hadzibabic}},\ }\bibfield
   {title} {\bibinfo {title} {Critical dynamics of spontaneous symmetry
  breaking in a homogeneous bose gas},\ }\href@noop {} {\bibfield  {journal}
  {\bibinfo  {journal} {Science}\ }\textbf {\bibinfo {volume} {347}},\ \bibinfo
  {pages} {167} (\bibinfo {year} {2015})}\BibitemShut {NoStop}%
\bibitem [{\citenamefont {Mahler}\ \emph {et~al.}(2019)\citenamefont {Mahler},
  \citenamefont {Goh}, \citenamefont {Tradonsky}, \citenamefont {Friesem},\
  and\ \citenamefont {Davidson}}]{Mahler19}%
  \BibitemOpen
  \bibfield  {author} {\bibinfo {author} {\bibfnamefont {S.}~\bibnamefont
  {Mahler}}, \bibinfo {author} {\bibfnamefont {M.}~\bibnamefont {Goh}},
  \bibinfo {author} {\bibfnamefont {C.}~\bibnamefont {Tradonsky}}, \bibinfo
  {author} {\bibfnamefont {A.~A.}\ \bibnamefont {Friesem}},\ and\ \bibinfo
  {author} {\bibfnamefont {N.}~\bibnamefont {Davidson}},\ }\href@noop {}
  {\bibinfo {title} {Improved phase locking of laser arrays with nonlinear
  coupling}} (\bibinfo {year} {2019}),\ \Eprint
  {https://arxiv.org/abs/1910.04430} {arXiv:1910.04430 [physics.optics]}
  \BibitemShut {NoStop}%
\bibitem [{\citenamefont {Nixon}\ \emph {et~al.}(2011)\citenamefont {Nixon},
  \citenamefont {Fridman}, \citenamefont {Friesem},\ and\ \citenamefont
  {Davidson}}]{Micha11}%
  \BibitemOpen
  \bibfield  {author} {\bibinfo {author} {\bibfnamefont {M.}~\bibnamefont
  {Nixon}}, \bibinfo {author} {\bibfnamefont {M.}~\bibnamefont {Fridman}},
  \bibinfo {author} {\bibfnamefont {A.~A.}\ \bibnamefont {Friesem}},\ and\
  \bibinfo {author} {\bibfnamefont {N.}~\bibnamefont {Davidson}},\ }\bibfield
  {title} {\bibinfo {title} {Enhanced coherence of weakly coupled lasers},\
  }\href@noop {} {\bibfield  {journal} {\bibinfo  {journal} {Opt.\ Lett.}\
  }\textbf {\bibinfo {volume} {36}},\ \bibinfo {pages} {1320} (\bibinfo {year}
  {2011})}\BibitemShut {NoStop}%
\bibitem [{\citenamefont {Pal}\ \emph {et~al.}(2015)\citenamefont {Pal},
  \citenamefont {Trandonsky}, \citenamefont {Chriki}, \citenamefont {Barach},
  \citenamefont {Friesem},\ and\ \citenamefont {Davidson}}]{Vishwa15}%
  \BibitemOpen
  \bibfield  {author} {\bibinfo {author} {\bibfnamefont {V.}~\bibnamefont
  {Pal}}, \bibinfo {author} {\bibfnamefont {C.}~\bibnamefont {Trandonsky}},
  \bibinfo {author} {\bibfnamefont {R.}~\bibnamefont {Chriki}}, \bibinfo
  {author} {\bibfnamefont {G.}~\bibnamefont {Barach}}, \bibinfo {author}
  {\bibfnamefont {A.~A.}\ \bibnamefont {Friesem}},\ and\ \bibinfo {author}
  {\bibfnamefont {N.}~\bibnamefont {Davidson}},\ }\bibfield  {title} {\bibinfo
  {title} {Phase locking of even and odd number of lasers on a ring geometry:
  effects of topological-charge},\ }\href@noop {} {\bibfield  {journal}
  {\bibinfo  {journal} {Opt.\ Express}\ }\textbf {\bibinfo {volume} {23}},\
  \bibinfo {pages} {13041} (\bibinfo {year} {2015})}\BibitemShut {NoStop}%
\bibitem [{\citenamefont {Chriki}\ \emph {et~al.}(2018)\citenamefont {Chriki},
  \citenamefont {Mahler}, \citenamefont {Tradonsky}, \citenamefont {Pal},
  \citenamefont {Friesem},\ and\ \citenamefont {Davidson}}]{Ronen2018}%
  \BibitemOpen
  \bibfield  {author} {\bibinfo {author} {\bibfnamefont {R.}~\bibnamefont
  {Chriki}}, \bibinfo {author} {\bibfnamefont {S.}~\bibnamefont {Mahler}},
  \bibinfo {author} {\bibfnamefont {C.}~\bibnamefont {Tradonsky}}, \bibinfo
  {author} {\bibfnamefont {V.}~\bibnamefont {Pal}}, \bibinfo {author}
  {\bibfnamefont {A.~A.}\ \bibnamefont {Friesem}},\ and\ \bibinfo {author}
  {\bibfnamefont {N.}~\bibnamefont {Davidson}},\ }\bibfield  {title} {\bibinfo
  {title} {Spatiotemporal supermodes: Rapid reduction of spatial coherence in
  highly multimode lasers},\ }\href@noop {} {\bibfield  {journal} {\bibinfo
  {journal} {Phys.\ Rev.\ A}\ }\textbf {\bibinfo {volume} {98}},\ \bibinfo
  {pages} {023812} (\bibinfo {year} {2018})}\BibitemShut {NoStop}%
\bibitem [{Sup()}]{Supplementary}%
  \BibitemOpen
  \bibinfo {note} {See Supplemental Material for additional technical details,
  experimental and simulated results.}\BibitemShut {Stop}%
\bibitem [{\citenamefont {Mehuys}\ \emph {et~al.}(1991)\citenamefont {Mehuys},
  \citenamefont {Streifer}, \citenamefont {Waarts},\ and\ \citenamefont
  {Welch}}]{Mehuys91}%
  \BibitemOpen
  \bibfield  {author} {\bibinfo {author} {\bibfnamefont {D.}~\bibnamefont
  {Mehuys}}, \bibinfo {author} {\bibfnamefont {W.}~\bibnamefont {Streifer}},
  \bibinfo {author} {\bibfnamefont {R.~G.}\ \bibnamefont {Waarts}},\ and\
  \bibinfo {author} {\bibfnamefont {D.~F.}\ \bibnamefont {Welch}},\ }\bibfield
  {title} {\bibinfo {title} {Modal analysis of linear talbot-cavity
  semiconductor lasers},\ }\href@noop {} {\bibfield  {journal} {\bibinfo
  {journal} {Opt.\ Lett.}\ }\textbf {\bibinfo {volume} {16}},\ \bibinfo {pages}
  {823} (\bibinfo {year} {1991})}\BibitemShut {NoStop}%
\bibitem [{\citenamefont {Trandonsky}\ \emph {et~al.}(2015)\citenamefont
  {Trandonsky}, \citenamefont {Nixon}, \citenamefont {Ronen}, \citenamefont
  {Pal}, \citenamefont {Chriki}, \citenamefont {Friesem},\ and\ \citenamefont
  {Davidson}}]{Chene15}%
  \BibitemOpen
  \bibfield  {author} {\bibinfo {author} {\bibfnamefont {C.}~\bibnamefont
  {Trandonsky}}, \bibinfo {author} {\bibfnamefont {M.}~\bibnamefont {Nixon}},
  \bibinfo {author} {\bibfnamefont {E.}~\bibnamefont {Ronen}}, \bibinfo
  {author} {\bibfnamefont {V.}~\bibnamefont {Pal}}, \bibinfo {author}
  {\bibfnamefont {R.}~\bibnamefont {Chriki}}, \bibinfo {author} {\bibfnamefont
  {A.~A.}\ \bibnamefont {Friesem}},\ and\ \bibinfo {author} {\bibfnamefont
  {N.}~\bibnamefont {Davidson}},\ }\bibfield  {title} {\bibinfo {title}
  {Conversion of out-of-phase to in-phase order in coupled laser arrays with
  second harmonics},\ }\href@noop {} {\bibfield  {journal} {\bibinfo  {journal}
  {Photon.\ Res.}\ }\textbf {\bibinfo {volume} {3}},\ \bibinfo {pages} {77}
  (\bibinfo {year} {2015})}\BibitemShut {NoStop}%
\bibitem [{\citenamefont {Tradonksy}\ \emph {et~al.}(2017)\citenamefont
  {Tradonksy}, \citenamefont {Pal}, \citenamefont {Chriki}, \citenamefont
  {Davidson},\ and\ \citenamefont {Friesem}}]{Chene17}%
  \BibitemOpen
  \bibfield  {author} {\bibinfo {author} {\bibfnamefont {C.}~\bibnamefont
  {Tradonksy}}, \bibinfo {author} {\bibfnamefont {V.}~\bibnamefont {Pal}},
  \bibinfo {author} {\bibfnamefont {R.}~\bibnamefont {Chriki}}, \bibinfo
  {author} {\bibfnamefont {N.}~\bibnamefont {Davidson}},\ and\ \bibinfo
  {author} {\bibfnamefont {A.~A.}\ \bibnamefont {Friesem}},\ }\bibfield
  {title} {\bibinfo {title} {Talbot diffraction and fourier filtering for phase
  locking an array of lasers},\ }\href@noop {} {\bibfield  {journal} {\bibinfo
  {journal} {Appl.\ Opt.}\ }\textbf {\bibinfo {volume} {56}},\ \bibinfo {pages}
  {A126} (\bibinfo {year} {2017})}\BibitemShut {NoStop}%
\bibitem [{Aco()}]{Acomment}%
  \BibitemOpen
  \bibinfo {note} {The vortex [anti-vortex] state is constructed by imposing
  the phases ($0, ~2\pi/3,~-2\pi/3$) [($0, ~-2\pi/3,~2\pi/3$)] on a certain
  triangle and then setting all other lasers in a unique manner to be separated
  by $\pm ~2\pi/3$ from their nearest neighbors.}\BibitemShut {Stop}%
\bibitem [{\citenamefont {Chalker}\ \emph {et~al.}(1992)\citenamefont
  {Chalker}, \citenamefont {Holdsworth},\ and\ \citenamefont
  {Shender}}]{Chalker92}%
  \BibitemOpen
  \bibfield  {author} {\bibinfo {author} {\bibfnamefont {J.~T.}\ \bibnamefont
  {Chalker}}, \bibinfo {author} {\bibfnamefont {P.~C.~W.}\ \bibnamefont
  {Holdsworth}},\ and\ \bibinfo {author} {\bibfnamefont {E.~F.}\ \bibnamefont
  {Shender}},\ }\bibfield  {title} {\bibinfo {title} {Hidden order in a
  frustrated system: Properties of the heisenberg kagome antiferromagnet},\
  }\href@noop {} {\bibfield  {journal} {\bibinfo  {journal} {Phys.\ Rev.\
  Lett.}\ }\textbf {\bibinfo {volume} {68}},\ \bibinfo {pages} {855} (\bibinfo
  {year} {1992})}\BibitemShut {NoStop}%
\bibitem [{\citenamefont {Moessner}\ and\ \citenamefont
  {Chalker}(1998)}]{Chalker98}%
  \BibitemOpen
  \bibfield  {author} {\bibinfo {author} {\bibfnamefont {R.}~\bibnamefont
  {Moessner}}\ and\ \bibinfo {author} {\bibfnamefont {J.~T.}\ \bibnamefont
  {Chalker}},\ }\bibfield  {title} {\bibinfo {title} {Low-temperature
  properties of classical geometrically frustrated antiferromagnets},\
  }\href@noop {} {\bibfield  {journal} {\bibinfo  {journal} {Phys.\ Rev.\ B}\
  }\textbf {\bibinfo {volume} {58}},\ \bibinfo {pages} {12049} (\bibinfo {year}
  {1998})}\BibitemShut {NoStop}%
\bibitem [{Bco()}]{Bcomment}%
  \BibitemOpen
  \bibinfo {note} {For NN, the probabilities for a state to have phase
  difference either $+ 2\pi/3$, or $-2\pi/3$ or $0$ are P(+)=1/2, P(-)=1/2 and
  P(0)=0. For NNN, the probabilities are P(+)=1/4, P(-)=1/4 and P(0)=1/2. For
  NNNN, the probabilities are P(+)=3/8, P(-)=3/8 and P(0)=1/4. In general, at a
  distance $n$ from the reference laser, the coherence varies as
  $1/2^{n}$.}\BibitemShut {Stop}%
\bibitem [{\citenamefont {Li}\ \emph {et~al.}(2017)\citenamefont {Li},
  \citenamefont {Lee}, \citenamefont {Huang}, \citenamefont {Burchesky},
  \citenamefont {Shteynas}, \citenamefont {Top}, \citenamefont {Jamison},\ and\
  \citenamefont {Keterle}}]{Keterle17}%
  \BibitemOpen
  \bibfield  {author} {\bibinfo {author} {\bibfnamefont {J.}~\bibnamefont
  {Li}}, \bibinfo {author} {\bibfnamefont {J.}~\bibnamefont {Lee}}, \bibinfo
  {author} {\bibfnamefont {W.}~\bibnamefont {Huang}}, \bibinfo {author}
  {\bibfnamefont {S.}~\bibnamefont {Burchesky}}, \bibinfo {author}
  {\bibfnamefont {B.}~\bibnamefont {Shteynas}}, \bibinfo {author}
  {\bibfnamefont {F.~C.}\ \bibnamefont {Top}}, \bibinfo {author} {\bibfnamefont
  {A.~O.}\ \bibnamefont {Jamison}},\ and\ \bibinfo {author} {\bibfnamefont
  {W.}~\bibnamefont {Keterle}},\ }\bibfield  {title} {\bibinfo {title} {A
  stripe phase with supersolid properties in spin–orbit-coupled
  bose–einstein condensates},\ }\href@noop {} {\bibfield  {journal} {\bibinfo
   {journal} {Nature}\ }\textbf {\bibinfo {volume} {543}},\ \bibinfo {pages}
  {91} (\bibinfo {year} {2017})}\BibitemShut {NoStop}%
\end{thebibliography}

\begin{thebibliography}{99}
\bibitem{MNixon13} M. Nixon, E. Ronen, A. A. Friesem, and N. Davidson, Observing geometric frustration with thousands of coupled lasers," Phys.\,Rev.\, Lett. \textbf{110}, 184102(2013).
\bibitem{TChene17} C. Tradonsky, V. Pal, R. Chriki, N. Davidson, and A. A. Friesem, ``Talbot diffraction and Fourier filtering for phase locking an array of lasers,'' Appl. Opt. \textbf{56}, A126 (2017).
\bibitem{PVishwa17}V. Pal, C. Tradonsky, R. Chriki, A. A. Friesem, and N. Davidson, ``Observing dissipative topological defects with coupled lasers," Phys.\,Rev.\,Lett.\,\textbf{119}, 013902 (2017).
\bibitem{Fox61} A. G. Fox and T. Li, ``Resonant modes in a maser interferometer,'' Bell System Technical Journal \textbf{40}, 453-488 (1961).
\bibitem{GS72} R. W. Gerchberg and W. O. Saxton, ``A practical algorithm for the determination of the phase from image and diffraction plane pictures,'' Optik \textbf{35}, 237-246 (1972).
\end{thebibliography}
\providecommand{\noopsort}[1]{}\providecommand{\singleletter}[1]{#1}%
\pagebreak
\widetext
\vskip0.5cm
\begin{center}
{\Large \bf Supplemental Material} 
\end{center}

\setcounter{equation}{0}
\setcounter{figure}{0}
\setcounter{table}{0}
\setcounter{page}{1}
\makeatletter
\renewcommand{\theequation}{S\arabic{equation}}
\renewcommand{\thefigure}{S\arabic{figure}}
\renewcommand{\bibnumfmt}[1]{[S#1]}
\renewcommand{\citenumfont}[1]{S#1}

\baselineskip=15pt

\setcounter{page}{1}
\vspace{4pt}
\section{Uncoupled lasers in square lattice}
\begin{figure}[htbp]
\centering
\includegraphics[width=100mm]{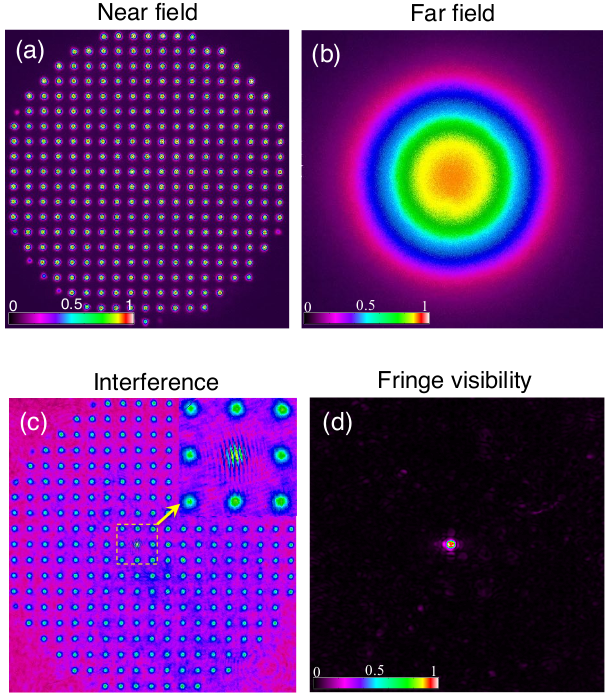}
\caption{Experimental results for the uncoupled lasers in a square lattice. (a) The near-field intensity pattern of lasers. (b) The far-field diffraction pattern of lasers. (c) The interference pattern when the output light from a single laser interferes with itself and with the light from all other lasers. (d) The coherence calculated from the measured interference pattern.}
\label{sqincoh}
\end{figure}

Here we show that without coupling the lasers in the lattice are independent from each other. The lattices of lasers are formed in a degenerate cavity, as shown in Fig.\,1(a) (shaded in yellow) in the manuscript. The results for the square lattice are shown in Fig.\,\ref{sqincoh}.

Figure\,\ref{sqincoh}(a) shows the near-field intensity pattern of the lasers arranged in the square lattice, where the output from each laser is a Gaussian TEM$_{00}$ mode. Fig.\,\ref{sqincoh}(b) shows the ensemble averaged far-field diffraction pattern that consists of a broad Gaussian distribution, which indicates that the lasers are independent from each other \cite{MNixon13}. We also measured the interference pattern using a Mach-Zehnder interferometer shown in Fig.\,1(a) (shaded in orange), where light from a single reference laser can interfere with itself and with the light from all other lasers. The results, shown in  Fig.\,\ref{sqincoh}(c) indicate that fringes only appear at one laser site (see inset in Fig.\,\ref{sqincoh}(c)), where the light from reference laser interferes only with itself. The corresponding coherence (fringe visibility) obtained by digital Fourier analysis is shown in Fig.\,\ref{sqincoh}(d) as a single spot, indicating that the selected laser is only coherent with itself and not with the other uncoupled lasers. These results confirm that the lasers in the array are independent and fully incoherent.

\section{Coupled lasers in square lattice}
\begin{figure}[htbp]
\centering
\includegraphics[width=110mm]{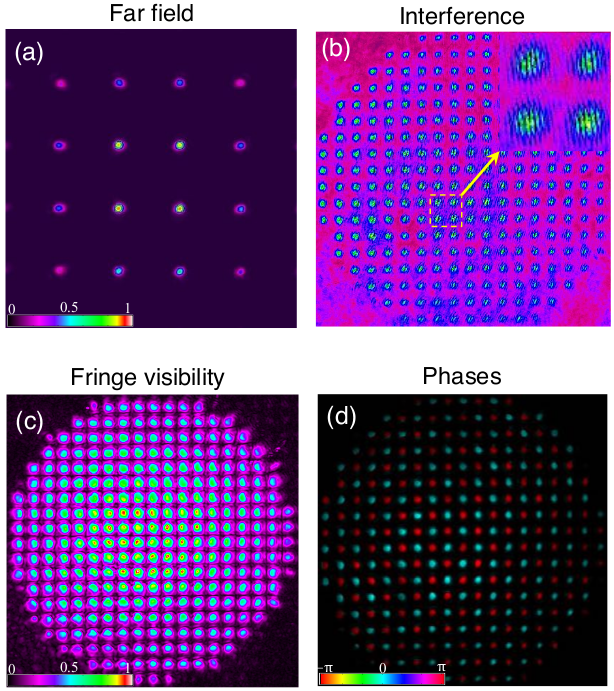}
\caption{Experimental results for the negatively coupled lasers in a square lattice. (a) The far-field diffraction pattern of laser lattice. (b) The interference pattern when the output light from a single reference laser interferes with itself and with the light from all other lasers. (c) The coherence calculated from the measured interference pattern. (d) The phases of the lasers calculated from the measured interference pattern.}
\label{sqcoh}
\end{figure}

Here we show the effect of coupling on the square lattice of lasers. We introduce coupling between the lasers by means of Talbot diffraction \cite{TChene17}, and then detect the effect on the output intensity distribution and coherence. The results are presented in Fig.\,\ref{sqcoh}.
Figure\,\ref{sqcoh}(a) shows the ensemble averaged far-field diffraction pattern which is comprised of sharp Bragg peaks with darkness in the center, indicating a long-range out-of-phase ordering. Figure\,\ref{sqcoh}(b) shows the interference pattern, where fringes were detected at all the laser sites, indicating a long range phase ordering. The analyzed coherence shown in Fig.\,\ref{sqcoh}(c) also indicates a long range phase ordering, where it barely decays with distance from the reference laser (center laser). Finally, Fig.\,\ref{sqcoh}(d)) shows the measured phases of the lasers (relative to the reference laser), confirming out-of-phase ordering throughout the lattice. All these results were obtained with ensemble averaging over 250 independent realizations, each corresponding to a different longitudinal mode. Accordingly, the detected long range out-of-phase ordering provides evidence that the entire ensemble of experiments (realizations) occupy the same non-degenerate ground state. This is equivalent to perfect anti-ferromagnetic spin ordering of XY spins.

\section{Triangular array}
\begin{figure}[htbp]
\centering
\includegraphics[width=80mm]{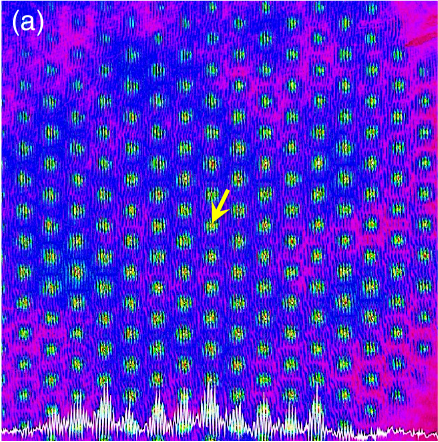}
\includegraphics[width=81mm]{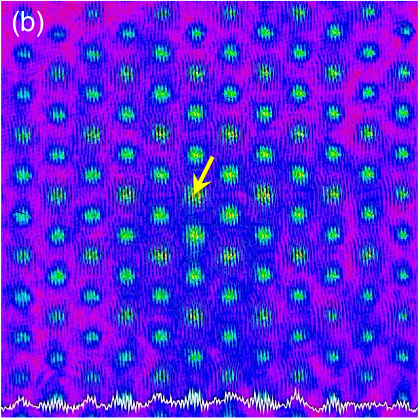}
\caption{\small Experimental interference patterns of coupled lasers in a triangular lattice. (a) Positively coupled lasers, as used to obtain the results Fig.\,2. (b) Negatively coupled lasers, as used to obtain the results of Fig.\,3. The yellow arrow denotes the location of the reference laser (same in other figures).}
\label{tricoh}
\end{figure}

Here we show the actual interference patterns from which we derived the coherence and phases in Figs.\, 2 and 3. Figure\,\ref{tricoh}(a) shows the interference pattern for the positively coupled lasers in a triangular lattice. It shows uniform coherence where the fringes barely disappear with distance from the reference laser. Figure\,\ref{tricoh}(b) shows the interference pattern of negatively coupled lasers, the interference pattern shows an oscillatory behavior, where coherence is maximal every third laser. 

\section{Kagome array}
\begin{figure}[htbp]
\centering
\includegraphics[width=83mm]{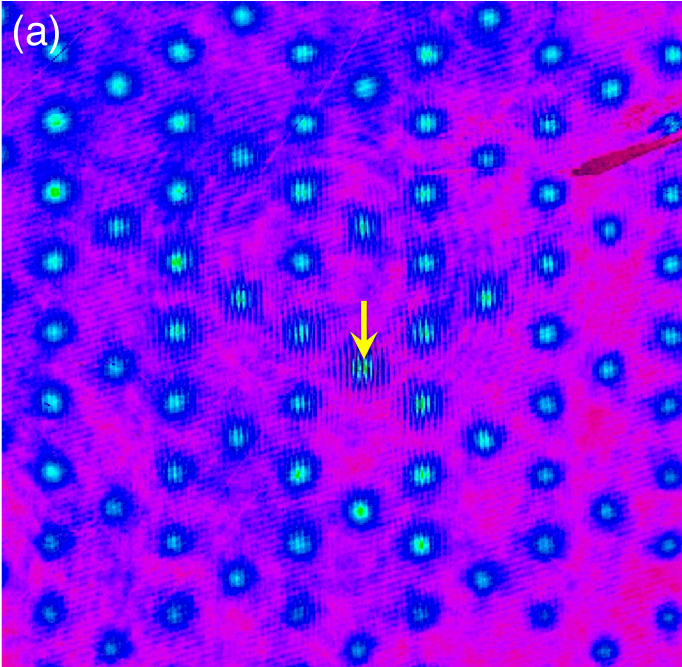}
\includegraphics[width=81mm]{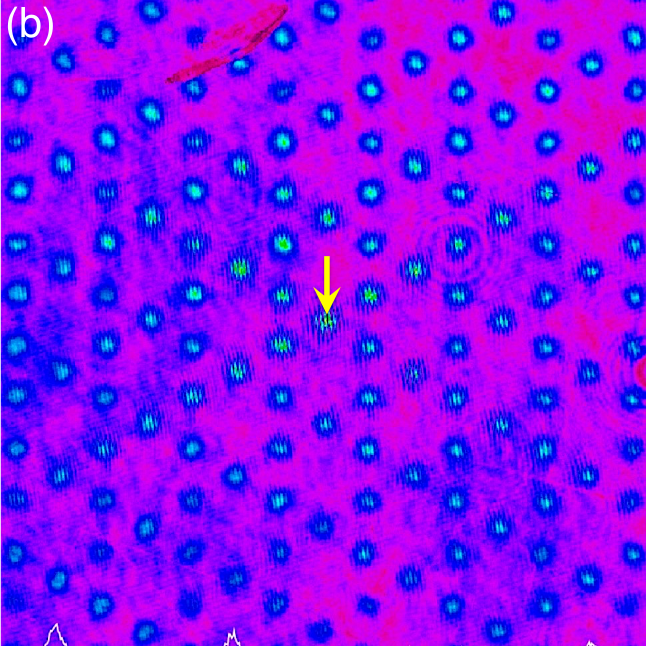}
\caption{\small Experimental interference patterns of negatively coupled lasers in a Kagome lattice. (a) With pure TEM$_{00}$ mode operation, as used to obtain the results Fig.\,4. (b) With no longer pure TEM$_{00}$ mode operation, as used to obtain the results of Fig.\,5.}
\label{Kagcoh}
\end{figure}

Here we show the actual interference patterns from which we derived the coherence and phases in  Figs.\,4 and 5.  Figure\,\ref{Kagcoh}(a) shows the interference pattern for the negatively coupled lasers in a Kagome lattice, where the operating mode of each laser is a pure TEM$_{00}$. As evident, the fringes are present only in a small central region, indicating a fast decay of coherence from the reference laser. Figure\,\ref{Kagcoh}(b) shows the interference pattern for the negatively coupled lasers, where the operating mode each laser is no longer a pure TEM$_{00}$ mode (consists of fine internal features). As evident, the fringes appear over long distances in certain directions, indicating a gradual decay of coherence from the reference laser along these certain directions.
\newpage
\section {Analytical results and numerical simulations}
\subsection{Analytical results}
\begin{figure}
\centering
\includegraphics[width=100mm]{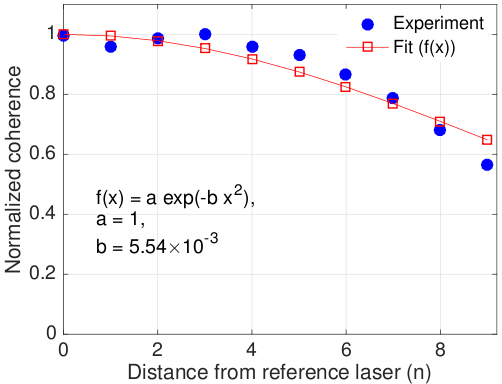}
\caption{\small Normalized coherence as a function of distance from the reference laser. The experimental visibility (blue dots) was fitted with a Gaussian decaying function (red line with squares) to extract the decay of coherence due to aberrations and noise.}
\label{Fit}
\end{figure}

Here we describe the method to calculate the decay of the coherence function shown in Fig.\,6(a) for the triangular lattice of lasers. We attribute the decay of the measured coherence to aberrations and noise in our experimental arrangement \cite{PVishwa17}. We first fitted the measured coherence of positively coupled lasers (in-phase ordered) with a Gaussian decay function
\begin{equation}
    f(x)=a~e^{-bx^{2}},
\end{equation}
where $a=1$, $b=5.54\times 10^{-3}$, and $x$ is the distance from the selected laser. The fitted curve is shown in Fig.\,\ref{Fit}. We multiplied the uniform coherence function ($f(x)=$ 1) of in-phase ordered triangular lattice and oscillatory coherence function of out-of-phase ordered triangular lattice ($f(x)=$ 1, 0.5, 0.5, 1, ...) by this Gaussian decaying function (Eq.\,(1)), yielding the analytical results in Fig.\,6(a).

\vspace{10pt}
Furthermore, the decay of the coherence function for the Kagome lattice shown in Fig.\,6(b) was analytically calculated by finding the probability distribution of states that have relative phase differences $\pm 2\pi/3$ or $0$ with respect to the reference laser, and found to decay exponentially as $1/2^{n}$, where $n$ denotes the distance from the reference laser.
\newpage
\subsection{Numerical simulations}
\begin{figure}
\centering
\includegraphics[width=150mm]{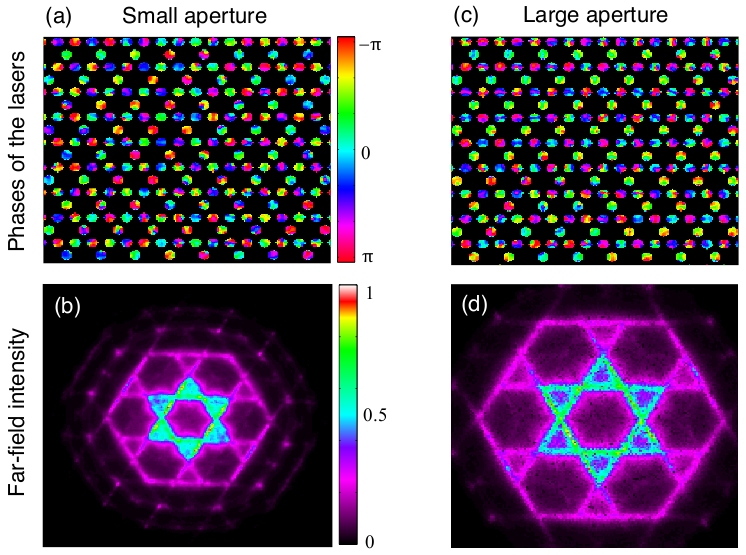}
\caption{\small Simulated phase and far-field diffraction pattern of negatively coupled lasers in a Kagome lattice for two different far-field aperture radii $R$. The top row shows the simulated phases of the lasers, and the bottom row shows the corresponding simulated far-field diffraction patterns.}
\label{Kagomelocked}
\end{figure}

Here, we describe the numerical simulations that was used to verify the experimental results of negatively coupled lasers in a Kagome lattice, shown in Figs.\,4 and 5.  The simulations were performed with an algorithm that combines the Fox-Li algorithm \cite{Fox61} and Gerchberg-Saxton algorithm \cite{GS72}. The parameters for the simulations were the same as those used in the experiment. The simulation results were averaged over $100$ realizations (corresponding to $100$ independent longitudinal modes) as to perform fair sampling. The simulated results are shown in Fig.\,\ref{Kagomelocked} for small and large far-field aperture radii $R$. The top row shows the simulated phases of the lasers, and the bottom row shows the corresponding far-field diffraction patterns. Note, the phases correspond to individual lasers, not the relative phase between the lasers. 

For small aperture $R=1$, Fig.\,\ref{Kagomelocked}(a) shows that the phase distribution in each laser is almost uniform (almost pure TEM$_{00}$ mode). The far-field diffraction pattern (Fig.\,\ref{Kagomelocked}(c)) shows large area Bragg lobes with diffusive lines similar to those in Fig.\,4(a), indicating lack of long-range phase ordering. For large aperture $R=1.2$, the phase distribution in each laser is mostly non uniform (no longer pure TEM$_{00}$ mode). The far-field diffraction pattern (Fig.\,\ref{Kagomelocked}(f)) shows sharp lines similar to those in Fig.\,5(a), indicating long-range phase ordering only along certain directions.
\newpage

\end{document}